# Electronic phase separation and emergence of a nondimerized insulating phase in $VO_2$ $(110)_R$ ultrathin films

S. Inoue[1], D. Shiga[1,2,*], R. Hayasaka[1], K. Ozawa[2], A. F. Santander-Syro[3], and H. Kumigashira[1,2]

[1] *Institute of Multidisciplinary Research for Advanced Materials (IMRAM), Tohoku University, Sendai 980–8577, Japan*

[2] *Photon Factory, Institute of Materials Structure Science, High Energy Accelerator Research Organization (KEK), Tsukuba 305–0801, Japan*

[3] *Université Paris-Saclay, CNRS, Institut des Sciences Moléculaires d'Orsay, 91405 Orsay, France*

**Abstract**

Using *in situ* photoemission spectroscopy and x-ray absorption spectroscopy, we investigated the thickness dependence of the electronic structure and V-V dimerization in $VO_2/TiO_2$ (110) ultrathin films, in which the one-dimensional V-V chains along the $c_R$ axis lie in the film plane. In $VO_2$ $(110)_R$ films, the reduction in dimensionality along the surface-normal direction is not expected to impose a geometric constraint on V-V dimerization, unlike in $VO_2$ $(001)_R$ films. Nevertheless, the characteristic spectral changes associated with the temperature-driven metal-insulator transition observed in thick films persist down to 1.5 nm, whereas at 1 nm an insulating electronic phase is observed without V-V dimerization. This behavior is highly similar to that reported for $VO_2$ $(001)_R$, suggesting that the enhancement of Mott instability resulting from reduced dimensionality is a common and essential driving force for the emergence of the nondimerized insulating phase in $VO_2$ ultrathin films. Meanwhile, unlike in $VO_2$ $(001)_R$, the nondimerized insulating phase in $VO_2$ $(110)_R$ coexists with the phase exhibiting the temperature-driven metal-insulator transition. Its fraction increases exponentially with decreasing thickness and becomes dominant at 1 nm. The corresponding effective critical thickness is estimated to be 2.2 nm. These results imply that the geometric orientation of the V-V chains dictates the spatial evolution of electronic phase separation via strain-mediated phase competition.

*Contact author: dshiga@tohoku.ac.jp

## I. INTRODUCTION

Vanadium dioxide ($VO_2$) is a representative strongly correlated oxide that exhibits a first-order metal-insulator transition (MIT) near room temperature, and its transition mechanism has been controversially discussed for over 60 years [1–10]. Across the MIT, both the electronic and crystal structures change simultaneously from the high-temperature rutile metallic (RM) phase to the low-temperature monoclinic insulating (MI) phase [2,3]. The MIT is also accompanied by a change in conductivity of several orders of magnitude [1,4]. Therefore, $VO_2$ has attracted broad interest from both fundamental and applied perspectives [11,12]. Moreover, the unusual MIT in $VO_2$, which originates from the interplay between lattice degrees of freedom and electron correlations, is regarded as a prototypical phenomenon for understanding the physics of strongly correlated oxides [4–7].

The structural phase transition in $VO_2$ is characterized by the tilting of V ions and V-V dimerization. In the low-temperature MI phase, V ions are collectively dimerized along the one-dimensional (1D) V-V chains oriented in the $c_R$ direction ($[001]_R$ direction), where $c_R$ denotes the $c$ axis of the rutile structure [3,9]. Although the MIT accompanied by such collective dimerization is reminiscent of the Peierls transition [13,14], many experimental and theoretical studies have demonstrated that electron correlations also play an essential role [4,6–8,15,16]. Accordingly, the MIT in $VO_2$ is now mainly understood as a cooperative Mott-Peierls transition [4,6–9]. To clarify the respective roles of these two cooperative instabilities, extensive efforts have been devoted to modulating their balance by various external stimuli, such as photoexcitation [17,18], strain control [19,20], and carrier injection through electrostatic gating or surface electron doping [21–25]. Nevertheless, how electron correlations and lattice instability each contribute to the MIT remains under debate.

Another promising approach to disentangle the roles of these two cooperative instabilities is to exploit surface-orientation dependence in epitaxial films. Surface-orientation-dependent properties of epitaxial $VO_2$ films have been extensively investigated [19,26–37]. In this context, the relative geometric arrangement between the confinement direction (film growth direction) and the V-V chain direction is expected to strongly influence the cooperative Mott-Peierls transition. In epitaxial films, the V-V dimerization axis can be oriented either in plane or out of plane with respect to both the surface and the interface simply by changing the substrate orientation. Such orientation control is a degree of freedom unique to epitaxial films and provides an effective means to examine the coupling between structural and electronic degrees of freedom in the MIT.

The influence of this geometric arrangement is expected to become particularly pronounced as the thickness is reduced toward the two-dimensional (2D) limit. For example, in $VO_2$ $(001)_R$ films, where the 1D V-V chains are oriented along the surface-normal direction, the reduction in thickness imposes a direct geometric constraint on the 1D V-V chains (corresponding to a reduction in the number of V-V dimers). In fact, it has been reported that a Mott insulating phase without dimerization emerges through the combined effects of the suppression of the Peierls instability caused by confinement effects on the 1D chains and the enhancement of electron correlations arising from reduced dimensionality [38]. Therefore, to separate the contributions of electron correlations and lattice instability, it is essential to investigate the effect of reduced dimensionality while excluding such direct geometric constraints.

From this viewpoint, $VO_2$ $(110)_R$ films provide an ideal counterpart. Because the V-V dimerization axis is oriented in plane (perpendicular to the confinement direction), reducing the film thickness does not impose a geometric constraint on the dimerization, unlike in the case of $VO_2$ $(001)_R$. Accordingly, $VO_2$ $(110)_R$ provides an excellent experimental platform for examining correlation effects induced by reduced dimensionality while minimizing the suppression of the Peierls instability, thereby allowing us to disentangle the respective roles of the two instabilities in the cooperative Mott-Peierls transition of $VO_2$.

Although several studies have addressed the electronic and structural properties of $VO_2$ films [19,28–30,33], systematic investigations correlating the electronic and crystal structures in the ultrathin regime below a few nanometers, where the MIT is expected to be substantially modified, remain very limited. Therefore, it is still unclear how the reduction in dimensionality along the film growth direction modifies the electronic phases of $VO_2$ under conditions where the V-V dimerization is not geometrically constrained.

Against this background, in this study, we aimed to clarify the thickness dependence of the electronic structure and V-V dimerization in $VO_2$ $(110)_R$ ultrathin films grown on $TiO_2$ (110) substrates. We prepared epitaxial $VO_2$ $(110)_R$ films with good crystallinity by pulsed-laser deposition (PLD) and systematically investigated the electronic structure and V-V dimerization in these dimensionality-controlled films using *in situ* photoemission spectroscopy (PES) and x-ray absorption spectroscopy (XAS). The spectroscopic results reveal three major findings. (1) The characteristic spectral changes associated with the temperature-driven MIT in thick films persist down to 1.5 nm. (2) In contrast, at 1 nm, an unusual phase with an insulating electronic structure but without V-V dimerization, namely, a nondimerized insulating (NI) phase, is observed. (3) In the intermediate-thickness regime, both the PES and XAS spectra are well described by linear combinations of the spectra characteristic of the temperature-driven MIT in the thick-film regime and those of the NI phase at 1 nm. The NI-phase fraction increases exponentially with decreasing film thickness, yielding an effective critical thickness of 2.2 nm. Aside from this distinct phase-separation behavior, the resulting electronic phase diagram is highly similar to that of $VO_2$ $(001)_R$. These results suggest that the emergence of an insulating phase without dimerization in $VO_2$ ultrathin films is primarily driven by the enhancement of Mott instability resulting from reduced dimensionality, irrespective of the relative geometric arrangement between the confinement direction and the 1D V-V chains. Meanwhile, the observed phase-separation behavior in $VO_2$ $(110)_R$ films implies that the geometric orientation of the V-V chains dictates the spatial evolution of electronic phase separation via strain-mediated phase competition.

## II. EXPERIMENT

Thickness-controlled $VO_2$ films were grown on 0.05 wt % Nb-doped rutile-$TiO_2$ (110) substrates by PLD. A sintered $V_2O_5$ pellet was used as the PLD ablation target. The deposition rate was estimated to be 0.02 nm $s^{-1}$ based on the film thickness determined by x-ray diffraction (XRD). The thickness was controlled in the range of 1–20 nm by changing the deposition time. During deposition, the substrate temperature and oxygen partial pressure were maintained at 400 °C and 10 mTorr, respectively.

After growth, the $VO_2$ $(110)_R$ films were transferred to the analysis chamber using a mobile vacuum suitcase under UHV at a pressure below $5 \times 10^{-10}$ Torr. This *in vacuo* transfer was critical for preserving the pristine film surfaces and preventing surface contamination and further oxidation [39–42]. PES measurements were carried out *in situ* using a VG-Scienta SES-2002 analyzer. The total energy resolutions at photon energies of 700 and 1200 eV were 120 and 200 meV, respectively. XAS measurements were also performed *in situ* with linearly polarized light by monitoring the sample drain current. For linear dichroism (LD) measurements of O *K*-edge XAS used to evaluate V-V dimerization, the spectra were acquired at $\theta = 0°$ and 60°, where $\theta$ denotes the angle between the $c_R$ axis (V-V dimerization direction) and the polarization vector **E**. The angle between the surface normal and the incident light was kept fixed during the measurements (see Fig. S1 in Supplemental Material [43]). The Fermi level ($E_F$) of each sample was determined by measuring a gold foil electrically connected to the sample.

The surface structure and cleanliness of the grown $VO_2$ films were confirmed by reflection high-energy electron diffraction and core-level PES, respectively. Detailed characterization of the films is presented in Supplemental Material [43]. The surface morphologies of the measured films were analyzed by atomic force microscopy in air, and the rms roughness values were 0.3 nm or less (Fig. S2). The crystal structure and epitaxial relationship were characterized by XRD (Fig. S3). The abruptness of the interface was also evaluated by core-level PES measurements, confirming a chemically abrupt interface (Fig. S4). The crystallinity and the surface and interface quality of the present films are comparable to those reported previously [19,33]. These results guarantee that the precondition for discussing the intrinsic thickness dependence is fulfilled, namely, that the films possess high crystallinity, smooth and clean surfaces, and chemically abrupt interfaces. The electrical resistivity was measured by the standard four-probe method.

## III. RESULTS

### A. Electronic phase diagram of $VO_2$ $(110)_R$ ultrathin films

First, we investigated the thickness ($t$) dependence of the transport properties of $VO_2$ $(110)_R$ ultrathin films. Figure 1(a) shows the temperature ($T$) dependence of resistivity ($\rho$) for $VO_2$ $(110)_R$ films grown on $TiO_2$ (110) substrates under identical growth conditions. For the thick 20-nm $VO_2$ film, a steep change in resistivity associated with the MIT and thermal hysteresis characteristic of a first-order phase transition are observed. The transition temperature ($T_{MIT}$) is determined to be 355 K, and the resistivity change $\rho$(300 K)/$\rho$(420 K) is approximately $5 \times 10^2$. These values are generally consistent with those previously reported for epitaxial $VO_2$ films grown on $TiO_2$ (110) substrates [19,32,33], indicating that the present films are of high quality and comparable to those investigated in previous studies. Taken together with the surface and interface characterizations presented in Supplemental Material [43], these results indicate that $t$ is the sole control parameter among the present samples. Meanwhile, compared with $VO_2\,(001)_R$ films [19,38,44], the present $VO_2\,(110)_R$ films exhibit $T_{MIT}$ values approximately 60–70 K higher and more gradual changes in the $\rho$-$T$ curves. These tendencies have already been reported to originate from differences in epitaxial strain between the two orientations [19,33].

With decreasing $t$, the MIT broadens, and $T_{MIT}$ apparently shifts to higher temperatures, reaching 364 K at $t$ = 4.5 nm. The magnitude of the resistivity changes across the MIT decreases with decreasing $t$, although a clear transition feature persists down to $t$ = 4.5 nm. However, the MIT disappears for $t \leq 3$ nm, where the $\rho$-$T$ curves exhibit no discernible kink and remain insulating over the entire measured temperature range. The results are summarized in the phase diagram shown in Fig. 1(b).

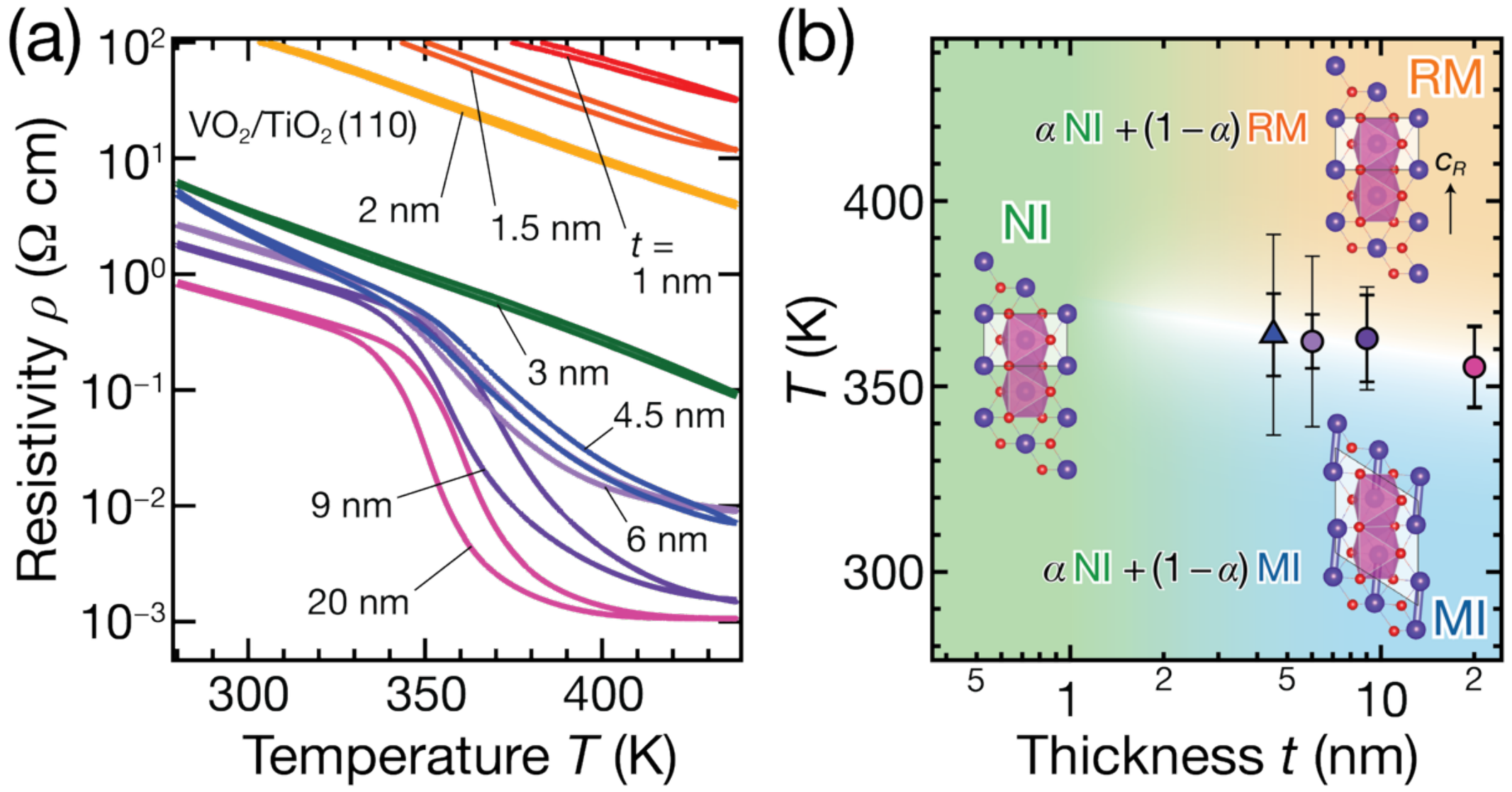

**FIG. 1.** (a) Temperature dependence of $\rho$ for $VO_2/TiO_2$ (110) films with $t$ ranging from 1 to 20 nm. (b) Electronic phase diagram of $VO_2$ $(110)_R$ ultrathin films as functions of $T$ and $t$. The circles indicate $T_{MIT}$ determined from the $\rho$-$T$ curves, and their colors correspond to those of the $\rho$-$T$ curves in (a). Here, $T_{MIT}$ is defined as the average of the inflection-point temperatures of the $\log_{10}\rho$-$T$ curves measured upon cooling and heating. The thick and thin vertical bars represent the hysteresis and transition widths in the $\rho$-$T$ curves [44], respectively. The MI, RM, and NI phases are assigned on the basis of the spectroscopic results (see Secs. III B–D). The intermediate-thickness regime (1 nm $< t \leq$ 9 nm) is denoted as phase separation. The high- and low-temperature states are described as $\alpha$NI + (1 − $\alpha$)RM and $\alpha$NI + (1 − $\alpha$)MI, respectively. Here, $\alpha$ denotes the NI-phase fraction deduced from the linear combination analysis of the PES and XAS spectra (see Sec. III D and Fig. 4 for details).

### B. Electronic structure determined by PES

Figure 2 shows the valence-band spectra for $VO_2$/Nb:$TiO_2$ (110) ultrathin films measured at $T$ = 420 K and 300 K, corresponding to the high-temperature (HT) and low-temperature (LT) states, respectively, at which the hysteresis loops in the $\rho$-$T$ curves are almost closed [Fig. 1(a)]. We carefully carried out temperature-dependent PES and XAS measurements while confirming that the spectral changes with temperature were saturated, namely, that hysteresis effects were no longer present. Furthermore, to avoid possible hysteresis effects, the sample temperature was maintained at 420 K for half an hour before the measurements, and subsequent measurements were performed only upon cooling [29,45]. We also confirmed that the spectral shapes remained unchanged even when the temperature was further increased in the HT state or decreased in the LT state, ensuring that the temperature-induced spectral changes were sufficiently saturated in both states. As shown in Fig. 2(a), the spectra of $VO_2$ $(110)_R$ are composed of two main features: structures derived from the O 2$p$ states at binding energies of 3–10 eV and peaks derived from the V 3$d$ states near $E_F$ [9,27,46–48]. As can be seen in Fig. 2, the Nb:$TiO_2$ substrate, which is a degenerate $n$-type semiconductor, has no significant occupied states within approximately 3 eV below $E_F$. Therefore, the electronic structures near $E_F$ of the $VO_2$ ultrathin films are not influenced by signals from the substrate, even in the ultrathin regime.

In the thick 20-nm film, the spectral changes characteristic of the temperature-driven MIT of $VO_2$ are clearly observed [9,27,29]. In the HT state, a sharp coherent peak at $E_F$ and a weak broad satellite structure around 1.2 eV are observed. On the other hand, in the LT state, a single peak appears around 0.8 eV, reflecting the formation of an energy gap at $E_F$. These characteristic spectral changes are in very good agreement with those previously reported for the RM and MI phases of $VO_2$ $(110)_R$ [27]. Furthermore, dramatic changes are also observed in the O 2*p* states across the temperature-driven MIT. These spectral changes originate from changes in the electronic and crystal structures accompanying the temperature-driven MIT of $VO_2$ [9,38,45].

With decreasing *t*, the characteristic spectral changes associated with the temperature-driven MIT remain almost unchanged down to $t = 4.5$ nm. Meanwhile, upon closer inspection of the spectral shape, we observe that the density of states represented by the coherent peak at $E_F$ in the HT state slightly decreases with decreasing *t* from 20 nm to 4.5 nm. This behavior corresponds to the fact that, in the $\rho$-$T$ curves shown in Fig. 1(a), $\rho$ gradually increases with decreasing *t*, while the characteristic MIT features are still maintained. When *t* is further reduced below 4.5 nm, the spectra show remarkable and systematic changes in the V 3*d* states near $E_F$. As can be seen from the expanded spectra in Fig. 2(b), at HT, the intensity of the coherent peak derived from V 3*d* states at $E_F$ decreases continuously and steeply with decreasing *t*. Importantly, the characteristic spectral changes with temperature remain discernible down to $t = 1.5$ nm. At $t = 1$ nm, a clear energy gap opens at $E_F$. This indicates that the HT electronic state progressively loses its metallic character with decreasing *t* and eventually becomes insulating in the ultrathin limit of $t = 1$ nm.

On the other hand, when the thickness dependence of the spectra at LT is examined, in contrast to the dramatic thickness-driven MIT observed at HT, it appears at first glance that there is no large change in the spectral shape, except for the decrease in the V 3*d* intensity accompanying the decrease in the number of V ions along the surface-normal direction of the film [Fig. 2(a)]. However, closer inspection of the line shape reveals that the 0.8-eV peak gradually shifts toward higher binding energy at $t \lesssim 3$ nm, eventually reaching 1.2 eV at $t = 1$ nm. Furthermore, the spectral change with temperature in the 1-nm film is barely visible except for thermal broadening. This result suggests that, in the ultrathin limit ($t = 1$ nm), both the HT and LT states become insulating and exhibit electronic structures different from that of the MI phase observed in the thick-film regime.

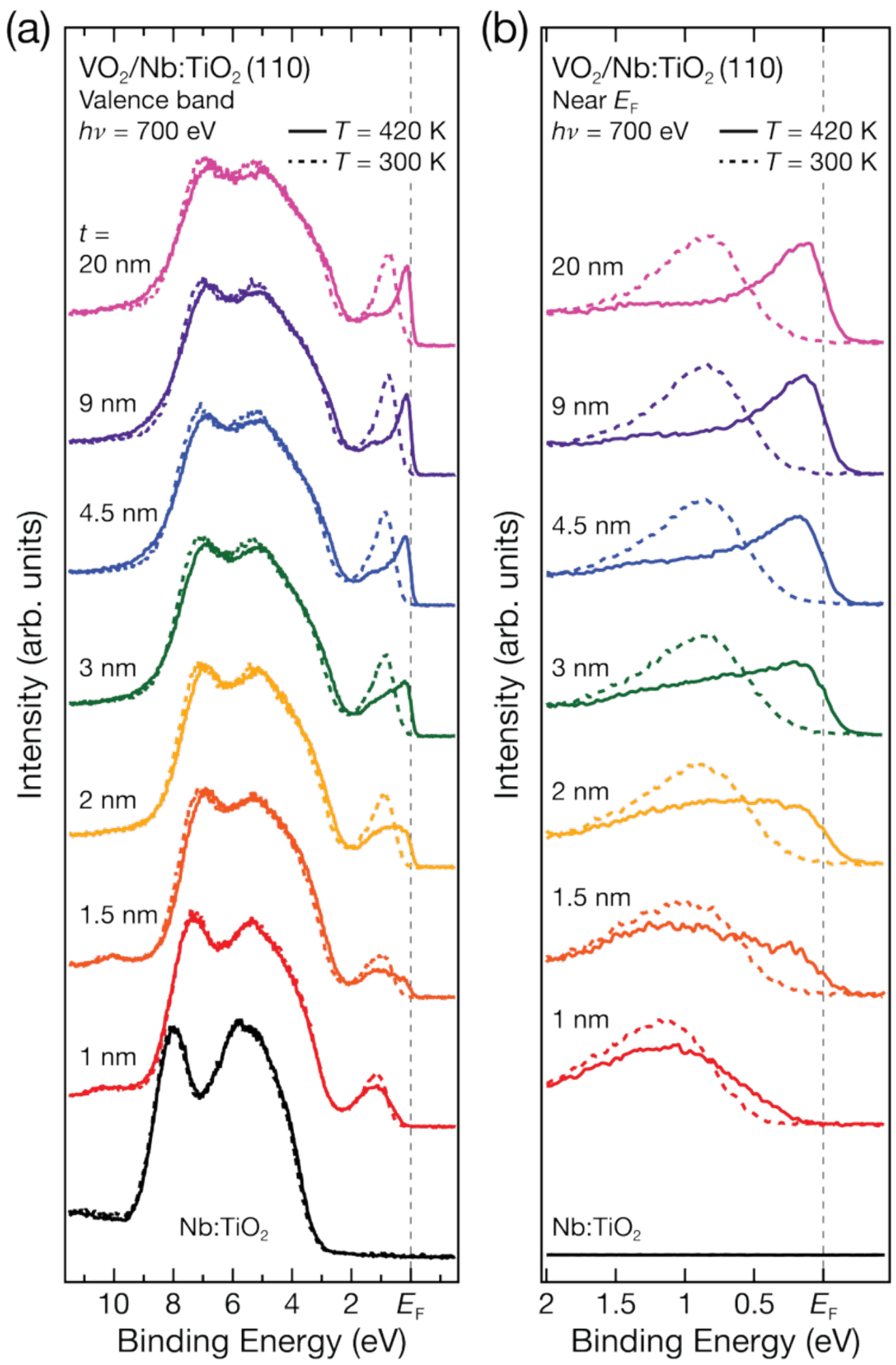

**FIG. 2.** (a) Valence-band spectra of $VO_2$/Nb:$TiO_2$ (110) ultrathin films with different $t$, measured at $T$ = 420 K (HT) and 300 K (LT). For the thick 20-nm film, these HT and LT states correspond to the RM and MI phases, respectively. (b) PES spectra near $E_F$ on an expanded energy scale. The colors of the spectra correspond to those used in Fig. 1.

### C. V-V dimerization studied by O *K*-edge XAS

Focusing on the O 2*p* states in the energy range of 3–10 eV [Fig. 2(a)], the dramatic changes across the temperature-driven MIT observed in thick films gradually weaken with decreasing $t$. At $t$ = 1 nm, no discernible spectral change with temperature is observed except for thermal broadening. For $VO_2$, the temperature-induced changes in the O 2*p* states are attributed to the structural changes accompanying the MIT [9], suggesting that the structural transition is suppressed with decreasing $t$ and eventually disappears in the ultrathin regime. This behavior, together with the remarkable changes in the V 3$d$ states near $E_F$ [Fig. 2(b)], strongly suggests the emergence of another insulating phase in the ultrathin regime [38].

To clarify the thickness dependence of this structural phase transition, namely, the V-V dimerization characteristic of $VO_2$, polarization-dependent O $K$-edge XAS measurements, which provide an appropriate indicator of this dimerization, were carried out as shown in Fig. 3. O $K$-edge XAS reflects the partial density of unoccupied O 2$p$ states hybridized with unoccupied V 3$d$ states and complements PES in examining the electronic structure of the conduction band. In the MI phase, the half-filled $d_{\|}$ state splits into the occupied bonding $d_{\|}$ state and the unoccupied antibonding $d_{\|}^*$ state owing to V-V dimerization, giving rise to an additional $d_{\|}^*$ peak in the XAS spectrum [3,9,32,49–51]. Furthermore, because of the strict dipole selection rule, this additional $d_{\|}^*$ state appears only when $\mathbf{E} \parallel c_R$. Based on the assignments in previous studies [9,32,50], in the MI phase, the corresponding peak appears at 530.6 eV in the spectrum with the $\mathbf{E} \parallel c_R$ configuration ($I_{\|}$), whereas it disappears in the RM phase. The assignment is further confirmed by the LD spectra: the additional $d_{\|}^*$ peak in the MI phase is absent in the spectrum corresponding to $\mathbf{E} \perp c_R$ ($I_{\perp}$; see Fig. S5 in Supplemental Material [43]). Therefore, the presence of the $d_{\|}^*$ peak in $I_{\|}$ can be used as a fingerprint of V-V dimerization in monoclinic $VO_2$ [3,9,50,51].

Figure 3(a) shows the thickness dependence of the O $K$-edge XAS spectra in the HT and LT states obtained with $\mathbf{E} \parallel c_R$ for the $VO_2$ $(110)_R$ ultrathin films. In the thick 20-nm film, the $d_{\|}^*$ peak is clearly observed at 530.6 eV at LT, whereas it disappears at HT. This indicates that the structural phase transition between the MI and RM phases occurs in the thick film. When $t$ is decreased, the $d_{\|}^*$ peak observed at LT gradually weakens starting at $t$ = 3 nm. As the film becomes even thinner, the intensity decreases steeply, and eventually the peak disappears at $t$ = 1 nm. This thickness dependence corresponds well to the suppression of the temperature-dependent changes in the O 2$p$ states in the valence-band spectra [Fig. 2(a)]. Furthermore, the leading-edge shift near the absorption edge [Fig. 3(a)] and the LD spectra [Fig. 3(b)] also show the same tendency. In fact, a clear temperature-induced leading-edge shift is observed near 529 eV for $t \geq 2$ nm, which is the counterpart of the energy-gap formation observed in the PES spectra in Fig. 2. Upon further decreasing $t$, the edge shift decreases steeply and eventually seems to disappear at $t$ = 1 nm.

These XAS observations indicate that V-V dimerization is no longer maintained at $t$ = 1 nm. Considering this together with the PES results in Fig. 2, it can be concluded that, in $VO_2\,(110)_R$, another insulating phase without V-V dimerization emerges in the ultrathin limit. At $t$ = 1 nm, this phase becomes dominant over the entire measured temperature range instead of the MI phase observed at LT in the thick-film regime. What should be emphasized here is that such an insulating phase emerges at a similar thickness scale irrespective of the orientation of the V-V dimerization axis with respect to the confinement direction [38]. This indicates that the emergence of the insulating phase in $VO_2$ ultrathin films is primarily governed by the enhancement of Mott instability resulting from reduced dimensionality rather than by the suppression of Peierls instability through geometric confinement.

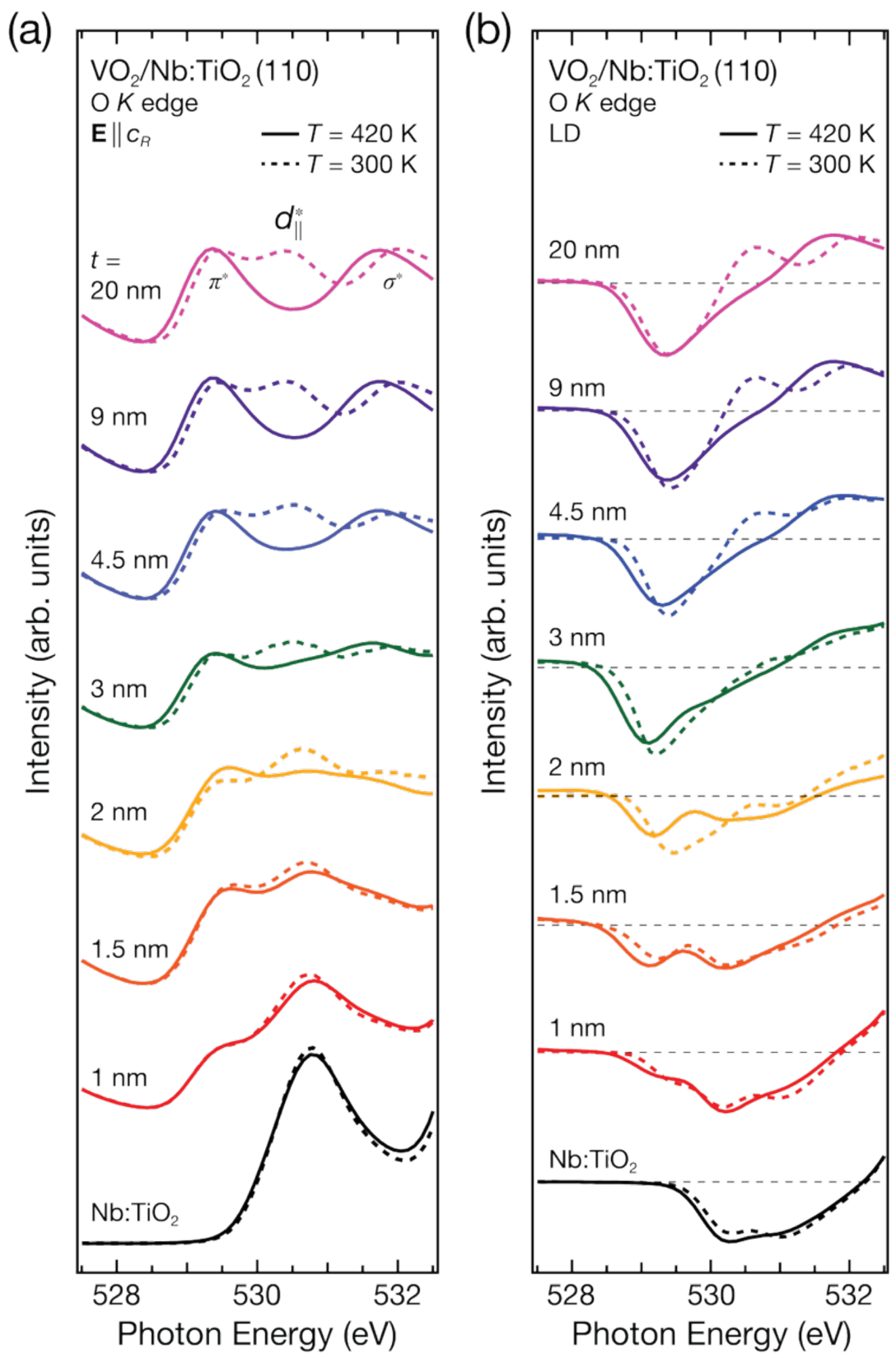

**FIG. 3.** (a) Thickness dependence of O *K*-edge XAS spectra of $VO_2$/Nb:$TiO_2$ (110) ultrathin films acquired in the **E** || $c_R$ geometry at $T = 420$ K (HT) and 300 K (LT). For the thick 20-nm film, the HT and LT states correspond to the RM and MI phases, respectively. (b) Corresponding LD spectra ($I_{\parallel} - I_{\perp}$). The colors of the spectra correspond to those used in Figs. 1 and 2.

### D. Phase coexistence analysis by linear combinations of PES and XAS spectra

Careful examination of the valence-band spectra (Fig. 2) and XAS spectra (Fig. 3) reveals a clear difference from $VO_2\,(001)_R$. In the intermediate-thickness regime (1 nm $< t \leq 9$ nm), both the HT and LT spectra appear to possess features characteristic of both the 20-nm film (temperature-driven MIT accompanied by V-V dimerization) and the 1-nm film (NI state over the entire measured temperature range). This suggests that some kind of electronic phase separation occurs during the thickness-dependent phase transition in $VO_2$ $(110)_R$. Therefore, to quantitatively examine the validity of this two-phase coexistence model, we performed the following linear-combination analysis.

Since no change in the spectral shape was observed even when the synchrotron-radiation spot was moved across the sample surface, it is considered that, if the two phases are spatially separated into domains, the domain sizes are sufficiently smaller than the soft-x-ray spot size used in the present experiment. In this case, the observed spectrum can be described as a linear combination of the spectra characteristic of the constituent phases. Therefore, in the analysis, we considered three reference phases: the RM and MI phases of the thick-film domain exhibiting the temperature-driven MIT accompanied by V-V dimerization and the NI phase that remains insulating over the entire measured temperature range. The spectra of the thick 20-nm film at HT and LT were used as the RM spectrum $I_{\mathrm{RM}}$ and the MI spectrum $I_{\mathrm{MI}}$, respectively, whereas the spectra of the 1-nm film at the corresponding temperatures were used as the NI spectra $I_{\mathrm{NI}}^{\mathrm{HT}}$ and $I_{\mathrm{NI}}^{\mathrm{LT}}$. Using these, the HT spectrum $I^{\mathrm{HT}}$ and LT spectrum $I^{\mathrm{LT}}$ at an arbitrary $t$ were fitted by the following equations [52]:

$$I^{\mathrm{HT}}(\alpha) = \alpha_{\mathrm{NI}}^{\mathrm{HT}} + (1-\alpha)I_{\mathrm{RM}}, \tag{1}$$

$$I^{\mathrm{LT}}(\alpha) = \alpha_{\mathrm{NI}}^{\mathrm{LT}} + (1-\alpha)I_{\mathrm{NI}}. \tag{2}$$

Here, $\alpha$ represents the NI-phase fraction obtained for each spectrum. $I_{\mathrm{RM}}$, $I_{\mathrm{MI}}$, $I_{\mathrm{NI}}^{\mathrm{HT}}$, and $I_{\mathrm{NI}}^{\mathrm{LT}}$ are each normalized, and $0 \leq \alpha \leq 1$.

Figures 4(a)–4(d) show the PES spectra near $E_F$ and O $K$-edge XAS spectra used for the linear-combination analysis, together with the fitting results using Eqs. (1) and (2). The contribution from the Nb:$TiO_2$ substrate was subtracted from the XAS spectra in Figs. 4(c) and 4(d). For both the HT and LT states, the PES and XAS spectra are well described by the linear-combination fits. This indicates that the NI and RM phases coexist at HT, whereas the NI and MI phases coexist at LT, and that $\alpha$ increases with decreasing $t$. The obtained $\alpha$ values are summarized in Fig. 4(e). Their thickness dependence is well described by an exponential function with an effective critical thickness, $\lambda$, of 2.2 nm.

The electronic phase diagram of $VO_2$ $(110)_R$ drawn on the basis of these comprehensive results is shown in Fig. 1(b). While the overall features are similar to those reported for $VO_2$ $(001)_R$ ultrathin films [38], a distinct difference is observed in the evolution of the electronic phases in the ultrathin regime. Specifically, whereas the entire $VO_2$ $(001)_R$ film undergoes an abrupt phase transition to the NI phase at $t$ = 1–1.5 nm, the NI phase in $VO_2$ $(110)_R$ already appears at relatively large thicknesses exceeding approximately 5 nm. Its fraction increases exponentially ($\lambda$ = 2.2 nm) with decreasing $t$, eventually becoming dominant at $t$ = 1–1.5 nm. It should be noted that the gradual increase in the NI-phase fraction may reconcile the discrepancy between the thickness scales inferred from the transport measurements ($t$ = 3–4.5 nm) [Fig. 1(a)] and the spectroscopic measurements ($t$ = 1–1.5 nm) from the viewpoint of percolation conduction [53–55].

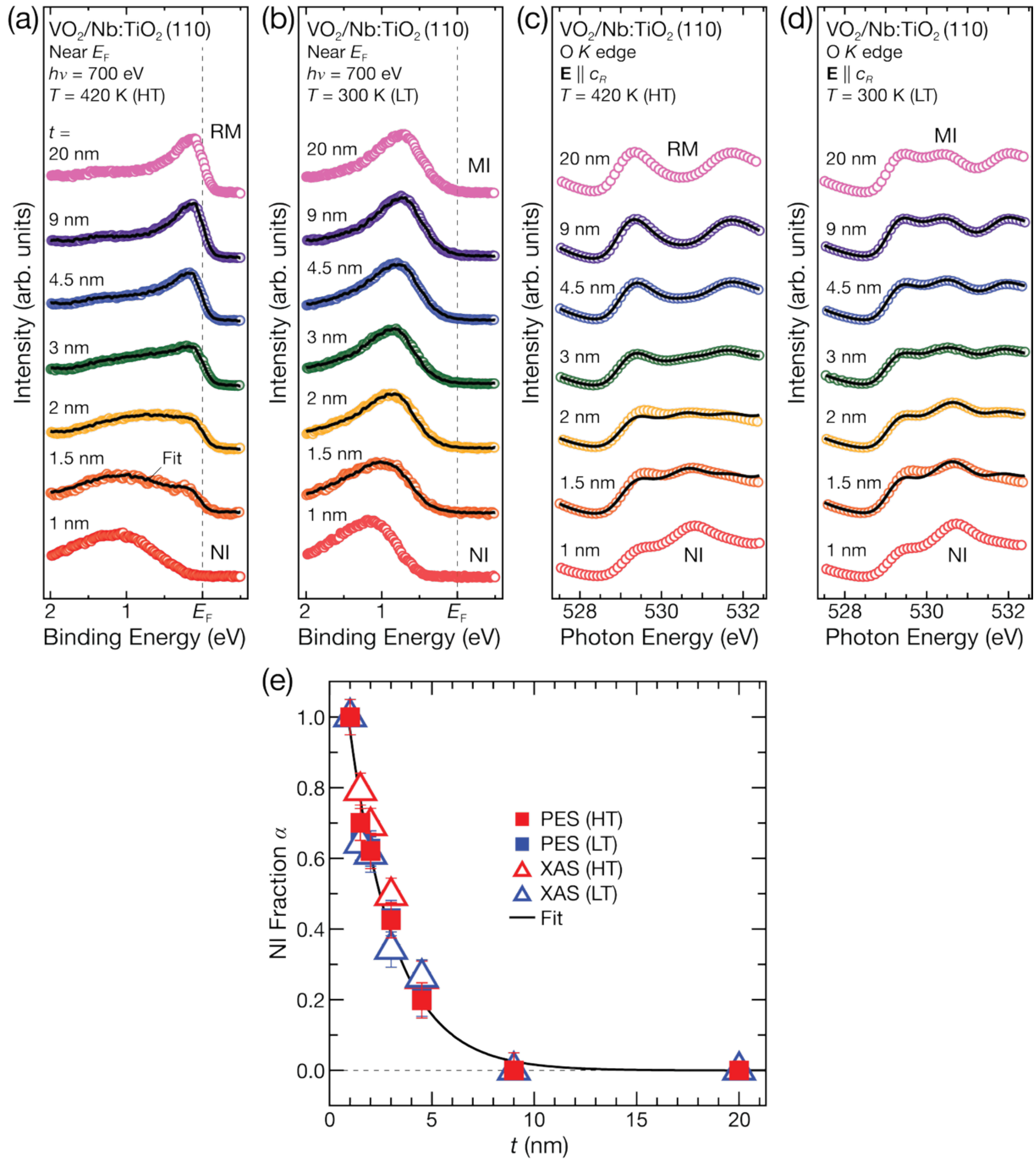

**FIG. 4.** Thickness dependence of valence-band spectra [(a) HT; (b) LT] and O *K*-edge XAS spectra [(c) HT; (d) LT] of $VO_2$/Nb:$TiO_2$ (110) ultrathin films. The PES spectra in (a) and (b) are identical to those shown in Fig. 2(b). The XAS spectra in (c) and (d) are shown after subtraction of the contribution from the Nb:$TiO_2$ substrate. The solid lines represent the linear-combination fits based on Eqs. (1) and (2). (e) Plot of $\alpha$ as a function of $t$. The $\alpha$ values obtained from (a)–(d) are fitted by an exponential function. The effective critical thickness is 2.2 nm.

## IV. DISCUSSION

The present systematic study of $VO_2$ $(110)_R$ films as a function of $t$ reveals two major findings. One is that, in the 2D limit ($t$ = 1 nm), the NI phase becomes dominant irrespective of the orientation of the V-V dimerization axis with respect to the confinement direction. The other is that, in $VO_2$ $(110)_R$, unlike in $VO_2$ $(001)_R$, the NI phase already coexists with the phase exhibiting the temperature-driven MIT at relatively large thicknesses, and its fraction continuously increases with decreasing $t$. In the following, we discuss these two results separately.

First, we discuss the stabilization mechanism of the NI phase in the 2D limit. In general, the Peierls instability in $VO_2$ is strongly connected with the 1D V-V chains along the $c_R$ direction [3,7,14]. Thus, in $VO_2$ $(001)_R$ ultrathin films, because the V-V dimerization axis is oriented along the out-of-plane direction, a reduction in $t$ acts as a geometric constraint that physically limits the number of V-V dimers [38]. In contrast, in $VO_2$ $(110)_R$, because the V-V dimerization axis is oriented in plane, the out-of-plane dimensional reduction does not impose a geometric constraint on V-V dimer formation. Nevertheless, a similar NI phase also appears in $VO_2$ $(110)_R$ in the 2D limit ($t$ = 1 nm). This experimental fact strongly indicates that, for the emergence of the NI phase in $VO_2$ ultrathin films, a reduction in the number of V atoms available for pairing due to direct geometric confinement is not an essential factor. Namely, the enhancement of Mott instability resulting from reduced dimensionality is considered to be a common and essential driving force for the emergence of the NI phase.

Next, we discuss the mechanism of the electronic phase separation observed with decreasing $t$. Although the origin of the orientation-dependent phase-separation behavior remains unclear, one possible origin is the competition among the elastic energy, the phase-boundary energy, and the free-energy difference. The balance among these energies may be specific to $VO_2/TiO_2$ (110), where the 1D V-V chains are oriented in plane [26,29,56]. Assuming a free-energy difference between the NI phase and the phase exhibiting the temperature-driven MIT and an elastic-energy gain arising from their coexistence, the energetic advantage of phase separation may depend strongly on the orientation of the V-V chains. In the $(001)_R$ geometry, where the V-V chains are oriented along the out-of-plane direction, phase coexistence does not provide an elastic-energy gain [57]. In contrast, in the $(110)_R$ geometry, where the V-V chains are oriented in plane, a phase-separated state may be stabilized through an elastic-energy gain. In fact, as shown in Fig.

1(a), the $\rho$-$T$ curves of $VO_2$ $(110)_R$ films show a more gradual transition than those of $VO_2$ $(001)_R$ films. Such broad transport behavior has been interpreted as a signature of strain-mediated electronic phase separation [26,29,30,58]. Therefore, compared with the $(001)_R$ orientation, the $(110)_R$ orientation provides an environment in which multiple phases compete and coexist more easily. This property may drive the phase coexistence with decreasing $t$ and the resulting percolative insulating behavior. Taken together, these results suggest that reduced dimensionality provides the essential driving force for the emergence of the NI phase, whereas the geometric orientation of the V-V chains dictates the spatial evolution of the phase separation through strain-mediated phase competition.

To fully elucidate the microscopic nature of the present phenomenon, further experimental and theoretical investigations will be desirable. Because this proposed scenario depends on a delicate energetic balance, identification of the crystal structure of the NI phase and detailed spatial information on the phase separation (domain distribution and size) are important. From this viewpoint, precise x-ray structural analysis of the NI phase and real-space observation of the phase separation using photoemission electron microscopy (PEEM), low-energy electron microscopy (LEEM), or mid-infrared scattering-type scanning near-field optical microscopy (s-SNOM) would provide effective approaches toward a comprehensive understanding of this phenomenon.

## V. CONCLUSION

The thickness dependence of the electronic structure and V-V dimerization in $VO_2$ $(110)_R$ ultrathin films grown on $TiO_2$ (110) substrates, in which the V-V dimerization axis lies in the film plane, was investigated by *in situ* PES and XAS. The main findings are summarized as follows. (1) The characteristic spectral changes associated with the temperature-driven MIT in thick films persist down to $t$ = 1.5 nm. (2) At $t$ = 1 nm, a nondimerized insulating (NI) phase is observed, as in $VO_2$ $(001)_R$ ultrathin films. (3) Both the PES and XAS spectra in the intermediate-thickness regime are well described by linear combinations of the spectra characteristic of the temperature-driven MIT in the thick-film regime and those of the NI phase at 1 nm. The NI-phase fraction increases exponentially with decreasing $t$, yielding an effective critical thickness of $\lambda$ = 2.2 nm.

These results show that the emergence of an insulating phase without dimerization in $VO_2$ ultrathin films is primarily driven by the enhancement of Mott instability resulting from reduced dimensionality, irrespective of the relative geometric arrangement between the confinement direction and the 1D V-V chains. Furthermore, the phase-separation behavior in $VO_2\,(110)_R$ films implies that the geometric orientation of the V-V chains dictates the spatial evolution of electronic phase separation via strain-mediated phase competition. These results provide important insights into the suppression and modulation of the cooperative Mott-Peierls transition under reduced dimensionality.

**ACKNOWLEDGMENTS**

The authors are very grateful to T. Yajima, T. Soma, and H. Tanaka for helpful discussions and thank H. Konaka, J. Yang, S. Watanabe, and Y. Masutake for their support during the experiments at the Photon Factory , KEK (KEK-PF). This work was supported by a Grant-in-Aid for Scientific Research (Nos. 20KK0117, 21K20498, 22H01948, 23K13664, 23K23216, 23H00263, and 25K01661) from the Japan Society for the Promotion of Science (JSPS), CREST (Grant No. JPMJCR2543) from the Japan Science and Technology Agency (JST), and MEXT Program: Data Creation and Utilization Type Material Research and Development Project (Grant No. JPMXP1122683430). S.I. acknowledges the financial support from the Division for Interdisciplinary Advanced Research and Education at Tohoku University. The experiments performed at KEK-PF were approved by the Program Advisory Committee of the Institute of Materials Structure Science, KEK (proposal Nos. 2022G675, 2024G656, 2021S2-002, and 2024S2-003).

## Supplemental Material

# Electronic phase separation and emergence of a nondimerized insulating phase in $VO_2$ $(110)_R$ ultrathin films

S. Inoue[1], D. Shiga[1,2,*], R. Hayasaka[1], K. Ozawa[2], A. F. Santander-Syro[3], and H. Kumigashira[1,2]

[1] *Institute of Multidisciplinary Research for Advanced Materials (IMRAM), Tohoku University, Sendai 980–8577, Japan*

[2] *Photon Factory, Institute of Materials Structure Science, High Energy Accelerator Research Organization (KEK), Tsukuba 305–0801, Japan*

[3] *Université Paris-Saclay, CNRS, Institut des Sciences Moléculaires d'Orsay, 91405 Orsay, France*

[*]Contact author: dshiga@tohoku.ac.jp

## Supplemental Note 1. Experimental geometry for polarization-dependent x-ray absorption measurements

Figure S1 depicts the experimental geometry for *in situ* polarization-dependent x-ray absorption spectroscopy (XAS) measurements, including the crystal axes of a $VO_2$ $(110)_R$ film and the polarization vector **E**. For linear dichroism measurements, the XAS spectra were acquired at $\theta = 0°$ ($\mathbf{E} \parallel c_R$) and 60°, where $\theta$ denotes the angle between the $c_R$ axis (V-V dimerization direction) and **E**. The angle between the surface normal and the incident light was kept fixed at 60° during the measurement, ensuring the same probing depth for the spectra acquired at the two $\theta$ values. At $\theta = 0°$, the measured spectrum corresponds to $I_{\parallel}$. The XAS spectrum with $\mathbf{E} \perp c_R$, $I_{\perp}$, is obtained from $I_{\perp} = (4/3)\,(I - I_{\parallel}/4)$, where $I$ is the spectrum measured at $\theta = 60°$.

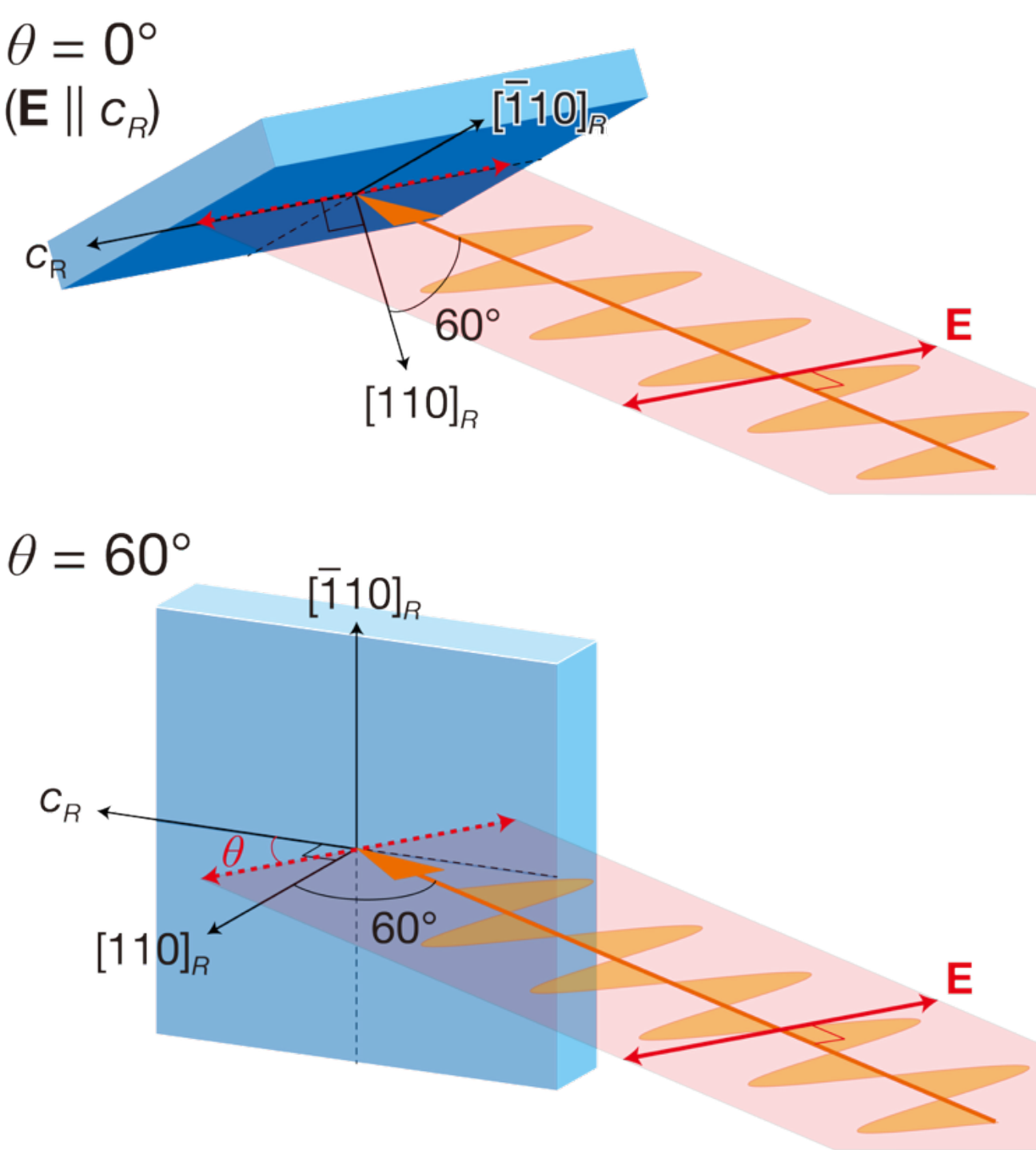

**FIG. S1.** Schematic of the experimental geometry of polarization-dependent XAS measurements of a $VO_2$ $(110)_R$ film at $\theta = 0°$ and 60°. The angle $\theta$ is defined as the angle between **E** and the $c_R$ axis. The $c_R$ axis corresponds to the $c$ axis of the rutile structure.

## Supplemental Note 2. Sample characterization

### A. Surface morphology

The atomically flat surfaces of all the measured $VO_2$/Nb:$TiO_2$ (110) films were confirmed by *ex situ* AFM, as shown in Fig. S2. An atomically flat step-and-terrace morphology inherited from the Nb:$TiO_2$ (110) substrate remains discernible up to $t$ = 9 nm, while smooth surface morphologies are maintained over the entire thickness range. The rms roughness, $R_{rms}$, estimated from the AFM images is 0.3 nm or less for all the films, i.e., comparable to or smaller than the separation between neighboring V-V chains along the surface-normal $[110]_R$ direction (approximately 0.318 nm), as determined by x-ray diffraction (see Supplemental Note 2B). These small $R_{rms}$ values indicate that local thickness inhomogeneity due to surface roughening is sufficiently suppressed even in the ultrathin regime, providing an essential precondition for discussing the intrinsic thickness dependence of the electronic and structural properties.

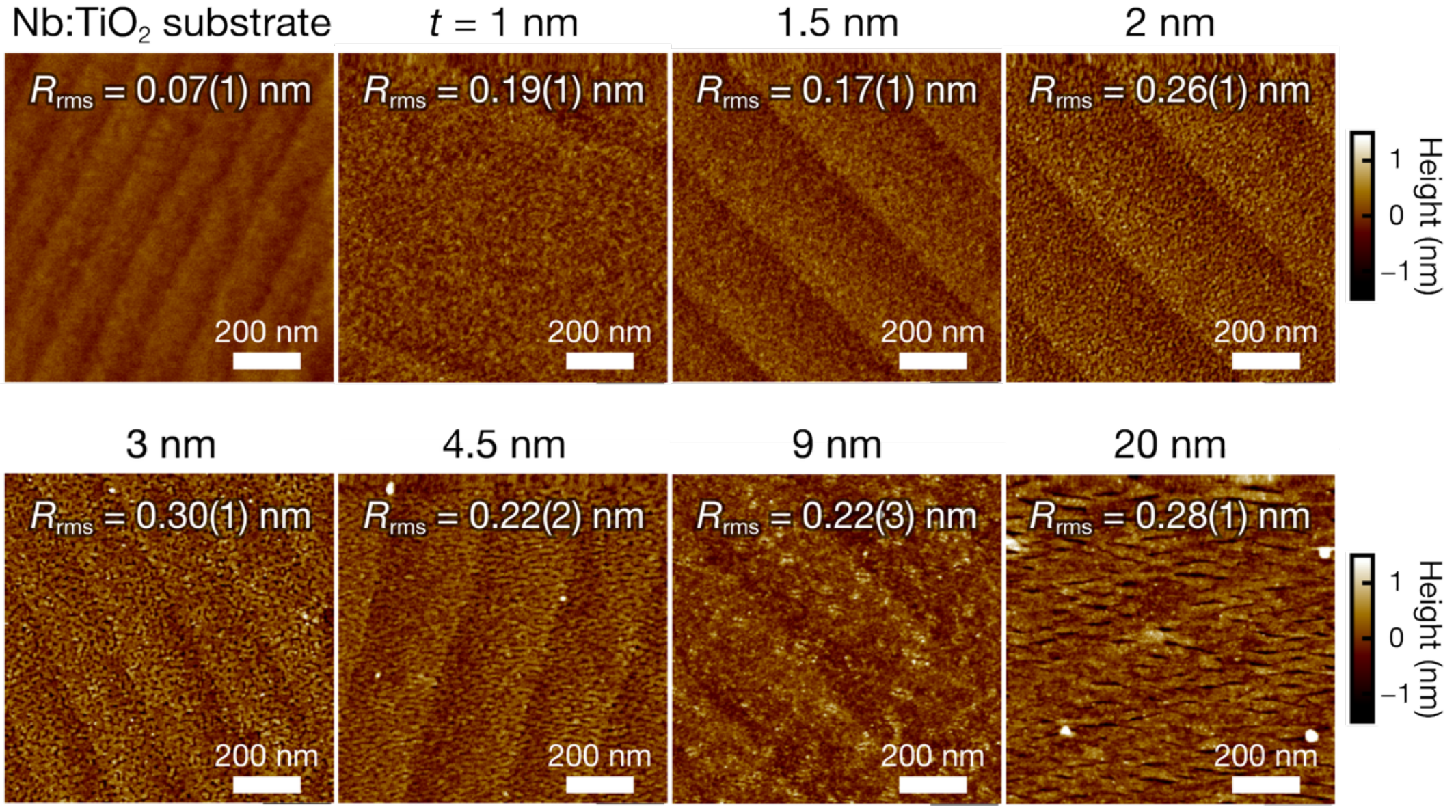

**FIG. S2.** AFM images of the measured $VO_2$ $(110)_R$ ultrathin films grown on Nb:$TiO_2$ (110) substrates with different $t$, together with that of the Nb:$TiO_2$ (110) substrate as a reference.

## B. Crystal structure

The crystal structure and out-of-plane epitaxial relationship of the $VO_2$/Nb:$TiO_2$ (110) ultrathin films were characterized by x-ray diffraction (XRD) measurements. Figure S3 shows the out-of-plane XRD patterns around the $TiO_2$ 110 and $VO_2$ $110_R$ reflections for the $VO_2$ $(110)_R$ films with different $t$. The observation of the $VO_2$ $110_R$ reflection together with the $TiO_2$ 110 reflection confirms the out-of-plane epitaxial relationship $VO_2$ $(110)_R$ || $TiO_2$ (110). For the thick 20-nm film, the distinct $VO_2$ $110_R$ reflection is accompanied by well-defined Laue fringes, indicating high crystalline quality, a uniform film thickness, and smooth surface and interface structures. This is consistent with the AFM results, which show $R_{\mathrm{rms}} \leq 0.3$ nm for all the measured films (see Supplemental Note 2A). The estimated out-of-plane spacing $d_{110_R}$ of the 20-nm film is approximately 0.318 nm, in good agreement with the value previously reported for epitaxial $VO_2$/$TiO_2$ (110) films [19,33]. The experimental XRD patterns are well reproduced by model simulations based on an epitaxial $VO_2$ $(110)_R$ film on a Nb:$TiO_2$ (110) substrate. With decreasing $t$, the diffraction intensity decreases and the Laue fringes become progressively less pronounced, becoming barely discernible for $t \leq 4.5$ nm. Importantly, this behavior is reproduced by the simulated XRD patterns by varying only $t$ while keeping $d_{110_R}$ fixed, indicating that the reduced visibility of the $VO_2$ $110_R$ reflection and Laue fringes in the ultrathin regime is attributable primarily to the small film thickness.

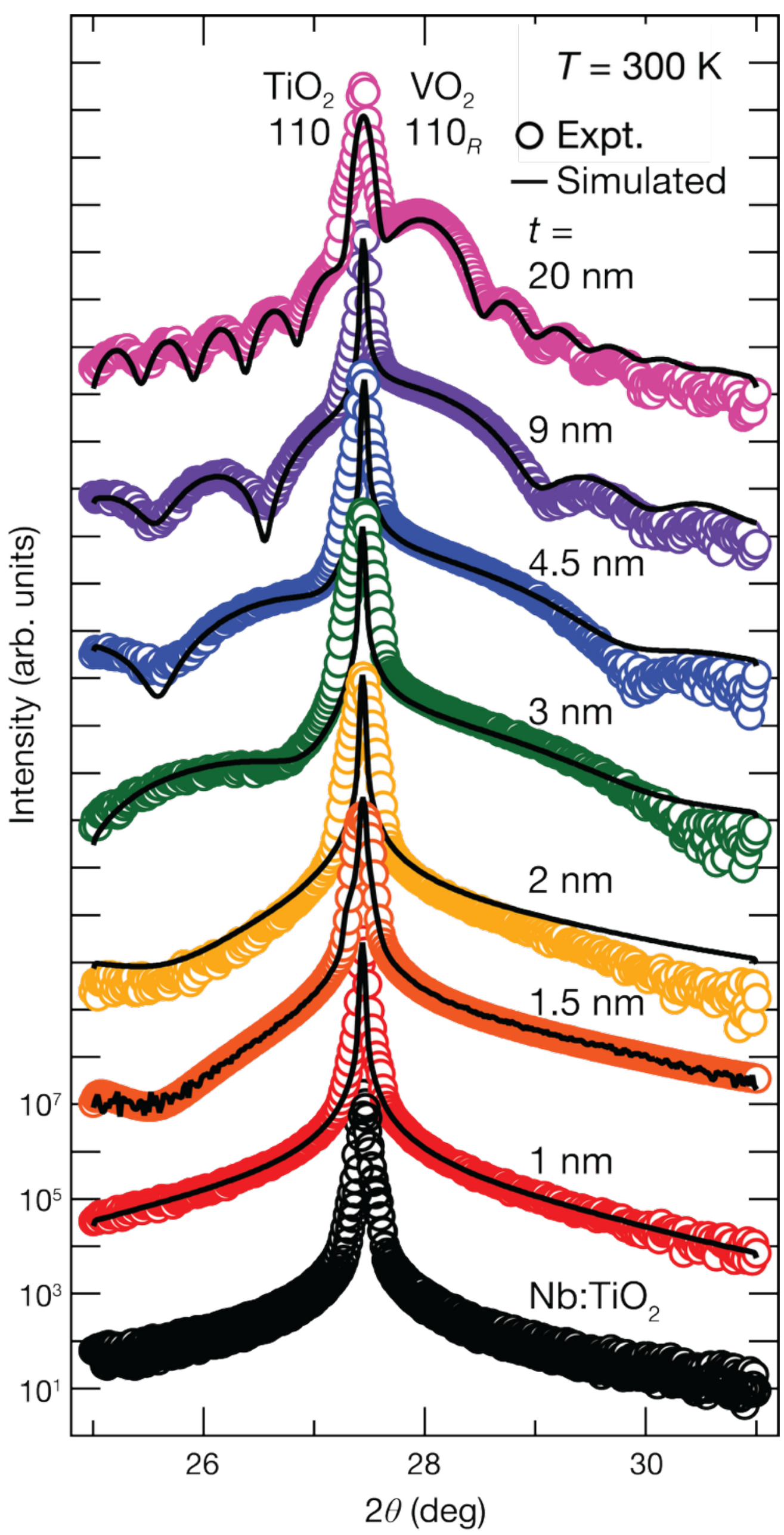

**FIG. S3.** Out-of-plane XRD patterns around the $TiO_2$ 110 and $VO_2$ $110_R$ reflections measured at room temperature for $VO_2\,(110)_R$ ultrathin films grown on Nb:$TiO_2$ (110) substrates with $t$ = 1–20 nm, together with that of the Nb:$TiO_2$ (110) substrate. The corresponding simulated XRD patterns based on the $VO_2\,(110)_R$/Nb:$TiO_2$ (110) heterostructure model are shown by the black curves.

**Supplemental Note 3. Core-level photoemission measurements**

Figure S4(a) shows the Ti $2p_{3/2}$ core-level spectra of $VO_2$/Nb:$TiO_2$ (110) films with different $t$, measured at 420 K, together with that of the Nb:$TiO_2$ (110) substrate as a reference. The spectra are normalized to the incident photon flux; hence, the intensity reduction with increasing $t$ reflects the attenuation of the Ti $2p$ photoelectron signal from the buried Nb:$TiO_2$ substrate by the $VO_2$ overlayer. The Ti $2p_{3/2}$ intensity decreases steeply with increasing $t$ and becomes nearly undetectable at $t = 9$ nm. To assess possible Ti interdiffusion at the interface, the relative Ti $2p_{3/2}$ intensity ($I_{\mathrm{Ti}}$) is plotted as a function of $t$ in the inset of Fig. S4(a) and fitted by the photoemission attenuation function $I_{\mathrm{Ti}}(t) = e^{-t/\lambda_{\mathrm{PES}}}$ [38], where $\lambda_{\mathrm{PES}}$ denotes the effective photoelectron attenuation length. The experimental data are well described by this function with $\lambda_{\mathrm{PES}} = 0.98(7)$ nm. This simple exponential attenuation is consistent with negligible Ti interdiffusion across the interface within the experimental sensitivity.

Figure S4(b) shows the Ti $2p_{3/2}$ spectra normalized to their peak intensities to visualize the thickness-dependent binding-energy shift. With increasing $t$, the Ti $2p_{3/2}$ peak shifts progressively toward lower binding energy and nearly saturates for $t \gtrsim 3$ nm [inset of Fig. S4(b)]. Judging from the saturation level, the magnitude of the band bending is estimated to be approximately 0.8 eV, comparable to that previously reported for $VO_2$/Nb:$TiO_2$ (001) [38]. This systematic binding-energy shift is attributed to band bending associated with the interfacial potential at the $VO_2$/$n$-type Nb:$TiO_2$ heterointerface.

To distinguish this electrostatic energy shift from possible changes in the chemical state of Ti, the spectra in Fig. S4(b) are aligned to their peak positions and compared in Fig. S4(c). The Ti $2p_{3/2}$ line shape remains essentially unchanged and retains the characteristic $Ti^{4+}$ line shape even at $t =$ 1 nm. No appreciable broadening or additional chemically shifted component is observed, indicating that the chemical environment of Ti is preserved near the interface. Together with the attenuation and band-bending behaviors in Figs. S4(a) and S4(b), respectively, these results provide consistent evidence for the formation of a well-defined, chemically abrupt $VO_2$/Nb:$TiO_2$ (110) interface.

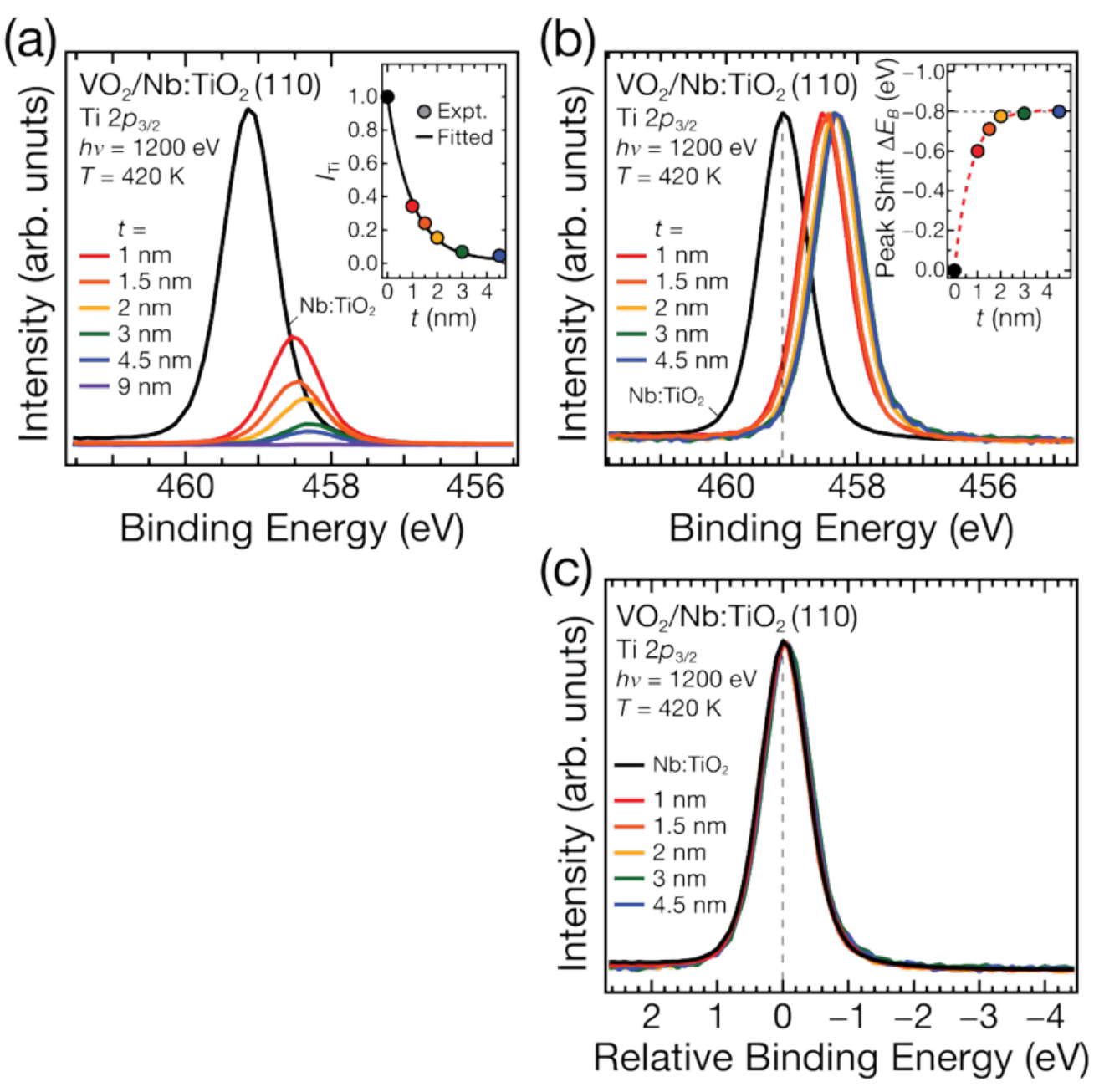

**FIG. S4.** (a) Thickness dependence of the Ti $2p_{3/2}$ core-level spectra of $VO_2$/Nb:$TiO_2$ (110) films measured at $h\nu = 1200$ eV and $T = 420$ K, together with that of the Nb:$TiO_2$ (110) substrate. The spectra are normalized to the incident photon flux. The inset shows $I_{Ti}$ as a function of $t$. The solid curve represents a fit using the photoelectron attenuation function, yielding $\lambda_{PES} = 0.98(7)$ nm. (b) Ti $2p_{3/2}$ spectra normalized to the peak intensities to highlight the binding-energy shift with $t$. The inset shows the peak shift $\Delta E_B = E_B(t) - E_B(0)$ as a function of $t$, where $E_B(0)$ denotes the $2p_{3/2}$ peak position of the Nb:$TiO_2$ (110) substrate; the red dashed curve is a guide to the eye. (c) Ti $2p_{3/2}$ spectra from (b) aligned to their peak positions to compare the line shapes. Panels (b) and (c) show the spectra for $t \leq 4.5$ nm, for which the Ti $2p_{3/2}$ signal remains sufficiently detectable.

## Supplemental Note 4. Polarization dependence of O *K*-edge XAS spectra

Figure S5 shows the polarization-dependent O *K*-edge XAS spectra of the $VO_2$/Nb:$TiO_2$ (110) films at $T = 300$ and 420 K. The $I_{\parallel}$ spectra ($\mathbf{E} \parallel c_R$; solid lines) are identical to those shown in Fig. 3(a) of the main text and are reproduced here for direct comparison with the $I_{\perp}$ spectra corresponding to $\mathbf{E} \perp c_R$ (dashed lines). These $I_{\parallel}$ and $I_{\perp}$ spectra were used to obtain the linear dichroism (LD) spectra, $I_{\parallel} - I_{\perp}$, shown in Fig. 3(b). For the thick 20-nm film at 300 K, the $d_{\parallel}^{*}$ peak around 530.6 eV appears in $I_{\parallel}$ but not in $I_{\perp}$; this $d_{\parallel}^{*}$-related polarization dependence gradually weakens with decreasing $t$ and is no longer discernible at $t = 1$ nm, consistent with the LD spectra in Fig. 3(b).

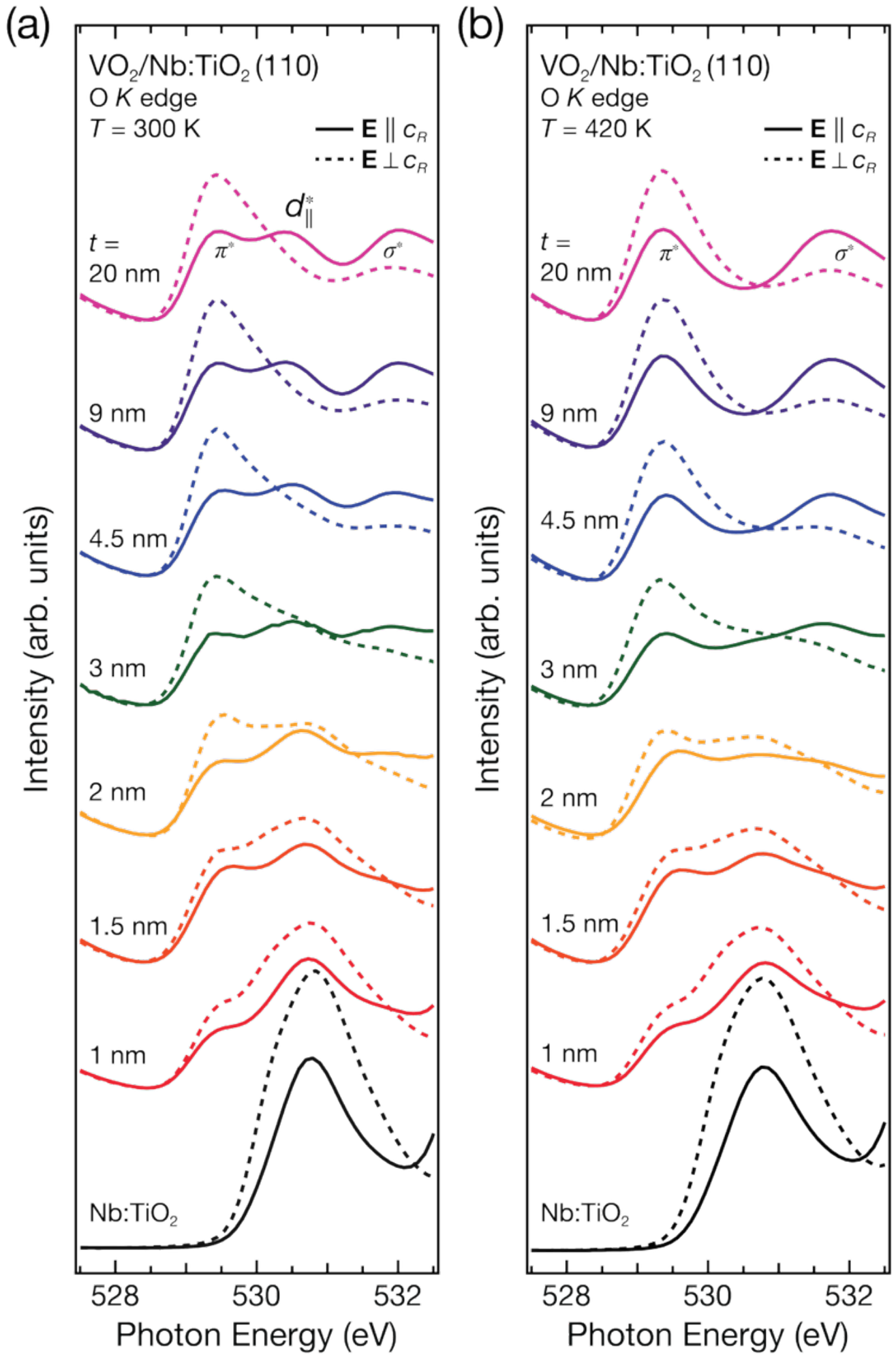

**FIG. S5.** Polarization-dependent O $K$-edge XAS spectra of $VO_2$/Nb:$TiO_2$ (110) films with different $t$ at (a) $T = 300$ K and (b) 420 K. The solid and dashed lines represent $I_{\parallel}$ ($\mathbf{E} \parallel c_R$) and $I_{\perp}$ (corresponding to $\mathbf{E} \perp c_R$), respectively.